\RequirePackage{fix-cm}
\documentclass{article}
\usepackage{graphicx}
\usepackage{mathptmx}      % use Times fonts if available on your TeX system
\usepackage{tikz}% we need them to plot commutative diagram
\usepackage{tikz-cd}
\usepackage{algorithm}
\usepackage{algpseudocode}
\usepackage{multirow}
\usepackage{longtable}
\usepackage[colorlinks,linkcolor=black,anchorcolor=black,citecolor=black,urlcolor=blue]{hyperref}
\usepackage{mathtools}
\usepackage{amssymb}
\usepackage{mciteplus}
\usepackage[superscript]{cite}
\usepackage{soul}
\usepackage{caption}
\usepackage{subcaption}
\usepackage{float}
\usepackage{booktabs}

\usepackage{geometry}
\usepackage[utf8]{inputenc}
\usepackage{algpseudocode}
\usepackage{algorithm}
\usepackage{hyperref}
\usepackage{mathrsfs}
\hypersetup{
    colorlinks=true,
    linkcolor=blue,
    filecolor=magenta,      
    urlcolor=magenta,
    citecolor=blue
    }

\title{Predicting Protein-Nucleic Acid Flexibility Using Persistent Sheaf Laplacians}
\author{Nicole Hayes$^1$, 
Ekaterina Merkurjev$^{1,2}$\footnote{Corresponding author,
	Email:  merkurje@msu.edu} ~ and 
 Guo-Wei Wei$^{1,3,4}$\\
$^1$ Department of Mathematics, \\
Michigan State University, MI 48824, USA.\\
$^2$ Department of Computational Mathematics, Science and Engineering\\
Michigan State University, MI 48824, USA.\\
$^3$ Department of Electrical and Computer Engineering,\\
Michigan State University, MI 48824, USA. \\
$^4$ Department of Biochemistry and Molecular Biology,\\
Michigan State University, MI 48824, USA. \\
}

\begin{document}

\date{}

\maketitle

\abstract{

Understanding the flexibility of protein–nucleic acid complexes, often characterized by atomic B-factors, is essential for elucidating their structure, dynamics,  and functions, such as reactivity and allosteric pathways. Traditional models such as Gaussian Network Models (GNM) and Elastic Network Models (ENM) often fall short in capturing multiscale interactions, especially in large or complex biomolecular systems. In this work, we apply the Persistent Sheaf Laplacian (PSL) framework for the B-factor prediction of protein–nucleic acid complexes. The PSL model integrates multiscale analysis, algebraic 
topology, combinatoric Laplacians, and sheaf theory for data representation. It reveals topological invariants in its harmonic spectra and captures the homotopic shape evolution of data with its non-harmonic spectra. Its localization enables accurate B-factor predictions. We benchmark our method on three diverse datasets, including protein–RNA and nucleic-acid-only structures, and demonstrate that PSL consistently outperforms existing models such as GNM and multiscale FRI (mFRI), achieving up to a 21\% improvement in Pearson correlation coefficient for B-factor prediction. These results highlight the robustness and adaptability of PSL in modeling complex biomolecular interactions and suggest its potential utility in broader applications such as mutation impact analysis and drug design.
}


%\newpage
 
\section{Introduction}
\label{sec:introduction}

Proteins and nucleic acids, including DNA and RNA, are some of the most essential building blocks of life. Proteins are involved in many vital processes, including cell signaling, gene regulation, transcription, and translation. Some of the key functions of proteins are binding (e.g., to DNA in DNA polymerase), catalysis (e.g., DNA polymerase catalyzing DNA replication), acting as molecular switches (e.g., GTPases catalyzing GTP hydrolysis, in turn switching "off" cellular processes), and providing structure to cells and organisms (e.g., actin, collagen, keratin, and silk) \cite{petsko2004protein}. Nucleic acids frequently act in conjunction with proteins to carry out important functions, such as gene expression and the storage and transmission of genetic information \cite{ollis1987structural, opron2016}. One of the most notable protein-nucleic acid complexes is the ribosome, mostly composed of RNA in addition to many smaller proteins, which synthesizes proteins and connects amino acids into polymer chains \cite{opron2016}. Another key complex is RNA polymerase, an enzyme that carries out transcription, the process of synthesizing RNA using a DNA sequence, or template \cite{petsko2004protein}.

Protein rigidity and flexibility are both crucial for protein structure and function \cite{Anfinsen1973, radivojac2004protein}. Protein rigidity gives rise to the three-dimensional (3D) structure of proteins, and this 3D structure determines protein functions \cite{petsko2004protein, branden2012introduction}. Protein flexibility plays a role in particular protein functions \cite{Frauenfelder1991} like folding and interactions with other molecules, including nucleic acids. Nucleic acid flexibility is crucial for other biological processes, such as the role of DNA and RNA flexibility in packing as well as interactions between nucleic acids and proteins \cite{hyeon2006size, opron2016}. While there has been extensive study of protein flexibility in recent decades \cite{ma2005usefulness,vihinen1994accuracy,jacobs2001protein,camps2009flexserv}, historically, much research on the flexibility of biomolecules primarily considered flexibility in the context of molecule motion and interaction. However, due to the discovery that proteins undergo thermal fluctuations even in neighborhoods of their native conformations (i.e., folded states), flexibility is now understood to be an intrinsic property of proteins \cite{McCammon1977, huber1983}.

There are multiple experimental approaches to measuring the flexibility of biomolecules, including X-ray crystallography, nuclear magnetic resonance (NMR), and single-molecule force experiments \cite{dudko2006intrinsic}. Protein flexibility, which can be measured by the B-factor, also called the Debye-Waller factor or temperature factor, is defined using the mean displacement of a scattering center during X-ray diffraction due to the thermal motion of atoms \cite{sun2019}. In addition to giving insight about the flexibility of atoms and amino acids in a protein structure, the B-factor also provides information about other protein functions, including the protein's thermal motion, activity, and structural stability \cite{yuan2005prediction}.

Due to the significant differences in proteins and nucleic acids, models attempting to analyze the flexibility of protein-nucleic acid complexes must account for this variability. For instance, amino acid residues and nucleotides---the building blocks of proteins and nucleic acids, respectively---have different length scales, and thus multiscale models such as PSL are advantageous for capturing this information. \cite{opron2016, hayes2025}. Many existing models for flexibility analysis utilize only one scale parameter, limiting their applicability on molecules with interactions over a wide range of scales. Notable examples include the anisotropic network model (ANM)\cite{Atilgan2001} and Gaussian network model (GNM)\cite{flory1976statistical, Bahar1997, haliloglu1997}, which are types of elastic network models (ENMs) and are some of the most popular methods for protein flexibility analysis \cite{xia2013multiscale, opron2014fast, xia2015multiscale}. ENMs treat the protein as a network, where the nodes are represented by the amino acid residues\cite{Atilgan2001,Bahar1998,Bahar1997,Hinsen1998,Li2002,Tama2001}. The first few eigenvalues of the network connectivity matrix reveal the long-time dynamics of proteins and can be used to predict B-factors. This process allows for the analysis of large proteins whose dynamics at large time scales would be intractable to traditional molecular dynamics (MD) simulations \cite{xia2013multiscale, Yang2008Coarse}.

The Gaussian network model typically outperforms the anisotropic network model in predicting B-factors \cite{opron2014fast, park2013coarse}, and GNM has also been shown to be about one order more efficient than other existing flexibility analysis models \cite{Yang2008Coarse}. Despite its advantages and popularity, GNM experiences difficulty in making predictions for many large biomolecules \cite{hinsen2008structural, kondrashov2007protein}. Park et al. \cite{park2013coarse} conducted extensive experiments applying GNM to predict B-factors on sets of relatively small, medium, and large protein structures. Although GNM achieved better results than normal mode analysis (NMA) on these data sets, it was unable to accurately predict B-factors for many structures.

Additional methods have emerged to overcome the disadvantages of GNM and ANM, including differential geometry analysis \cite{feng2025multiscale} as well as multiscale GNM (mGNM) and multiscale ANM (mANM) methods, which significantly improve protein B-factor prediction with respect to traditional GNM and ANM \cite{xia2015multiscale}. In particular, traditional GNM uses a single cutoff distance, limiting its predictive ability for molecules with interactions at multiple length scales, thereby motivating the mGNM and mANM methods. Another particularly successful method is the flexibility-rigidity index (FRI) \cite{xia2013multiscale}, which is built on the theory of continuum elasticity with atomic rigidity (CEWAR). FRI is a structure-based approach that relies on two assumptions: that protein functions are entirely determined by a protein's structure and environment, and that protein structure is determined by a protein's interactions \cite{opron2016}. These assumptions allow FRI to bypass the Hamiltonian interaction matrix used in ENMs, leading to significantly lower computational complexity. Adaptations such as fast FRI (fFRI) \cite{opron2014fast} have further streamlined the process. Additionally, multiscale FRI (mFRI) has been shown to improve the performance of FRI on certain challenging protein structures for GNM, again largely due to the single cutoff distance employed by GNM \cite{opron2015}. Both GNM \cite{yang2006} and FRI\cite{opron2016} (including mFRI) have also been used to predict the flexibility of protein-nucleic acid complexes, with two-kernel mFRI demonstrating marked improvement over both GNM and FRI \cite{opron2016}. More recently, an FRI method has been utilized for chromosome flexibility analysis, slightly improving predictive accuracy and significantly improving computational efficiency compared to GNM \cite{peng2021flexibility}.

Topological data analysis (TDA) has also been used as a technique for biomolecular study, and persistent homology has emerged as a particularly useful topological tool \cite{carlsson2009topology,edelsbrunner2008persistent}. TDA has been widely applied to molecular sciences \cite{wee2025review}. However, persistent homology has several drawbacks, including its inability to capture non-topological information \cite{wang2020persistent}. To address these drawbacks, persistent topological Laplacians (PTLs), also simply called persistent Laplacians, were introduced by Wei and his coworkers in 2019 \cite{wang2020persistent, chen2021evolutionary} as a new approach to integrate the topological and geometric knowledge gained from persistent homology and multiscale graphs, respectively. Many other PTLs have since been developed, including the persistent path Laplacian \cite{wang2023persistent}, persistent directed graph Laplacian \cite{jones2025persistent}, persistent hyperdigraph Laplacian \cite{wei2025persistent2}, and persistent sheaf Laplacian (PSL)\cite{wei2025persistent}. Like many TDA methods, most of these algorithms are global, making them ill-suited for flexibility analysis, which requires information about individual atom sites in a biomolecule \cite{bramer2020atom}. However, PSLs are capable of generating local information, thereby providing features for individual atoms in a protein and enabling the prediction of flexibility at specific sites. Other local persistent homology and persistent spectral methods have been used for RNA data analysis, including RNA flexibility prediction \cite{xia2023persistent, pun2020weighted}. In addition to encoding local information, PSLs also retain the benefits of other persistent topological methods by capturing geometric and non-topological information. For more detail about recent advances in TDA, we refer the reader to \cite{su2025topological}. %a comprehensive review \cite{su2025topological}. 

Recently, we have demonstrated the success of the persistent sheaf Laplacian (PSL) model in predicting protein B-factors\cite{hayes2025}. While many existing topological approaches to molecular biology produce information about a molecule as a whole, the PSL model enables atom-specific feature generation, supporting its use for protein flexibility analysis. Additionally, due to its use of cellular sheaves, the PSL model allows for the inclusion of non-spatial information in addition to the topological information inherent to many TDA methods, which contributes to its predictive ability.

In the present work, we extend the application of the PSL model from proteins to protein-nucleic acid complexes. Other models, such as multiscale FRI (mFRI), have similarly demonstrated success in both protein and protein-nucleic acid flexibility analysis. Thus, the multiscale nature of the PSL model, also a feature of persistent homology methods and other persistent topological methods, supports this extension.
Furthermore, the PSL method demonstrates marked improvement over GNM on a benchmark data set, achieving a 21\% increase in the average Pearson correlation coefficient compared to GNM for one representative model. In addition to its improvement over GNM, the PSL model also performs well compared to a two-kernel mFRI method from the literature \cite{opron2016}. One benchmark data set contains a subset of nucleic-acid-only structures, on which the PSL model also excelled compared to mFRI. This promising performance on varied structures demonstrates the strengths of the PSL model and suggests potential utility for other vital and complex molecules with interactions at multiple scales.

The remainder of this manuscript is organized as follows: Section \ref{sec:results} presents the results of the present work on data sets from the literature, Section \ref{sec:discussion} provides discussion of our results as well as a few case studies, and Section \ref{sec:methods} reviews persistent sheaf Laplacian theory and describes the method of PSL feature generation used in our experiments. Specifically, Section \ref{subsec:data-sets} introduces the data sets used for benchmarking the PSL model. Section \ref{subsec:parameters} reviews the parameters used in generating the PSL features for our model. Section \ref{subsec:64-203-results} results of the PSL model on two protein-nucleic acid data sets, including comparisons of the PSL method to existing results. Section \ref{subsec:126-results} presents new results for a set of protein-RNA structures, Section \ref{subsec:case-studies} contains a few selected protein-nucleic acid complex case studies to illustrate our model's advantages.

\section{Results}
\label{sec:results}

\subsection{Data Sets and Coarse-Grained Models}
\label{subsec:data-sets}

Given the demonstrated success of the PSL model for protein flexibility analysis \cite{hayes2025}, the present work extends this analysis to complexes consisting of both proteins and nucleic acids. Accordingly, the experiments in this section were performed on three data sets of protein-nucleic acid structures. The first data set, introduced by Yang et al. \cite{yang2006}, contains 64 structures, 19 of which consist of only nucleic acids (no amino acids) \cite{opron2016}. The second data set of 203 high-resolution protein-nucleic acid complexes was introduced by Opron et al. \cite{opron2016}. The third data set of specifically protein-RNA structures was previously used by Liu et al. \cite{liu2025} and Harini et al. \cite{harini2025} in analysis of binding affinity changes upon mutation. For PDB structures containing residues with alternate locations, we removed the lower-occupancy atomic coordinates in preprocessing, as outlined by Opron et al. \cite{opron2016}.

In order to effectively perform flexibility analysis on these data sets, we must utilize coarse-grained representations of the structures. In our earlier paper on protein flexibility analysis using the PSL model \cite{hayes2025}, we used a typical coarse-grained representation for proteins consisting of only the $C_{\alpha}$ atom from each residue. Similarly, Yang et al. \cite{yang2006} proposed three coarse-grained representations for nucleic acids, denoted M1, M2, and M3.

\begin{figure}[ht]
    \centering
    \includegraphics[width=0.7\textwidth]{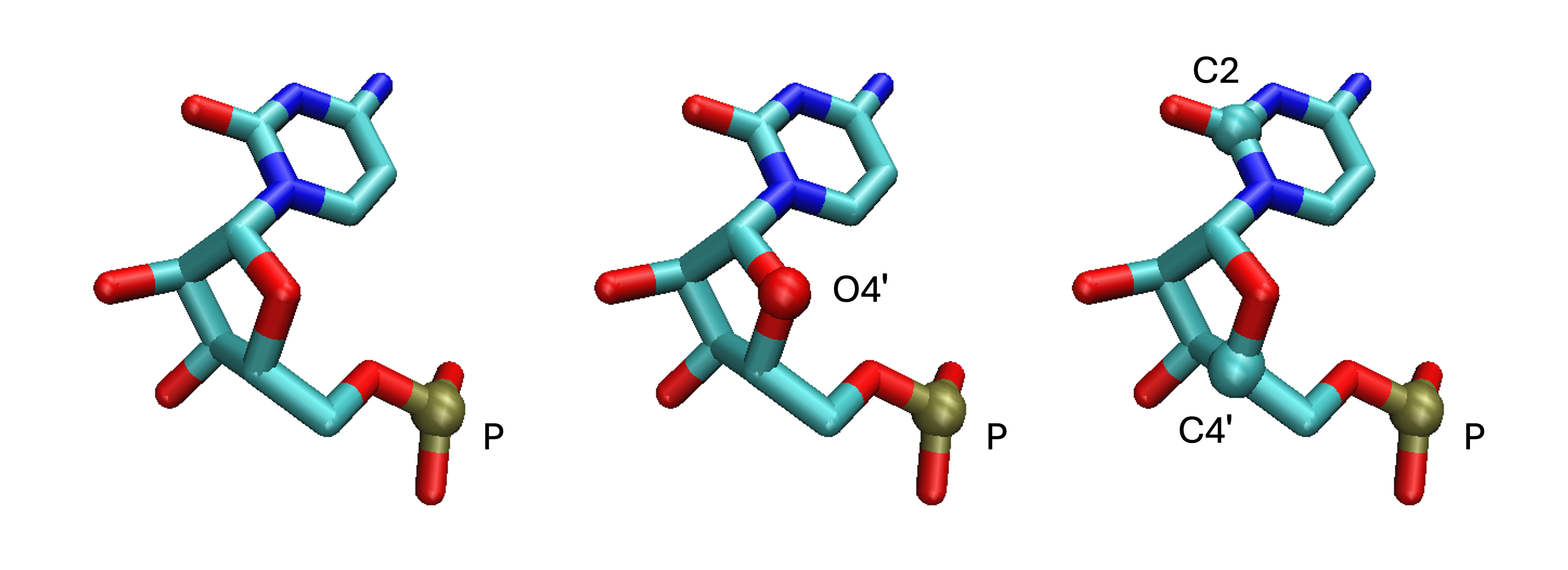}
    \caption{Visualization of the three coarse-grained models used in this work using the Visual Molecular Dynamics (VMD) software \cite{humphrey1996vmd}. The figure illustrates models M1, M2, and M3, from left to right, for a single residue of nucleic acid 2TRA. Each representation depicts the atoms included in that model, and atoms are colored by name. The backbone P atom is included in each model, with models M2 and M3 including one and two additional atoms per nucleotide, respectively.}
    \label{fig:coarse-grained-models}
\end{figure}

The M1 model consists of backbone P atoms for nucleotides and $C_{\alpha}$ atoms for proteins (i.e., one atom per nucleotide). The M2 model includes the atoms from M1 as well as sugar O4' atoms (i.e., two atoms per nucleotide). The M3 model contains the atoms from M1 in addition to sugar C4' atoms and base C2 atoms (i.e., three atoms per nucleotide) \cite{opron2016, yang2006}. Our experiments in the present paper utilize these three coarse-grained representations, and we compare our results to those of Opron et al. \cite{opron2016} and Yang et al. \cite{yang2006} using these same representations. Figure \ref{fig:coarse-grained-models} provides a visualization of the three coarse-grained models, indicating which atoms are included in each representation.

In addition to the data sets above, for which benchmark GNM results \cite{yang2006}, flexibility-rigidity index (FRI) results, and multiscale FRI (mFRI) results \cite{opron2016} exist in the literature, we also consider a data set of 141 protein-RNA complexes, which does not have existing B-factor prediction results. This data set was obtained from Liu et al. \cite{liu2025} and originally introduced by Harini et al. \cite{harini2025}. Both of these studies concerned experiments on 710 mutations of these complexes. To generate a data set for our own experiments, we extracted the unique PDB IDs of the complexes from the data set of 710 mutations and performed our analysis using the PDB files of these complexes. (Note that in the original paper \cite{harini2025}, it is stated that the 710 mutations originate from 134 protein-RNA complexes. However, there are 141 unique PDB IDs in the data set of 710 mutations. Thus, the present work refers to the set of 141 protein-RNA complexes.)

In our B-factor experiments, we excluded 15 of the 141 protein-RNA complexes due to unrealistic B-factors, resulting in a set of 126 structures. Specifically, PDB IDs 1AUD, 2JPP, 2KFY, 2KG0, 2KXN, 2LEB, 2LEC, 2LI8, 2M8D, 2MJH, 2MXY, 2RU3, 5M8I were excluded due to all atomic B-factors being equal to zero. The other two excluded structures were 2ERR (all B-factors were equal to 10.00) and 3J5S (all B-factors were equal to 1.00). We preprocessed the remaining 126 complexes as above to create M1, M2, and M3 models for each complex.

\subsection{Parameters}
\label{subsec:parameters}

For all sets of results in this work, we used filtration parameters of 6\r{A}, 12\r{A}, and 18\r{A} for the PSL model. Specifically, for each filtration radius, we constructed a 0th persistent sheaf Laplacian matrix $L_0$ and computed its eigenvalues. To generate PSL features for each radius, we then computed the maximum, minimum, mean, and median of the nonzero eigenvalues of $L_0$ and counted the number of its zero-valued eigenvalues. In addition to these five features for each of our three filtration radii, we also included the standard deviation of the nonzero eigenvalues of $L_0$ for 18\r{A}, resulting in 16 total features for downstream machine learning tasks. For the experimental B-factor predictions in this section, we performed linear regression on each data set as a whole using the PSL features.

\subsection{Results on 64 and 203 Structures}
\label{subsec:64-203-results}

In this section, we report our results for the PSL model for B-factor prediction of the protein-nucleic acid complexes in the data sets of 64 and 203 structures, respectively. The results for the data set of 64 complexes are displayed in Table \ref{64-results}, and the results for the 203-structure data set are shown in Table \ref{203-results}. Additionally, the results on nucleic-acid-only structures from the set of 64 complexes can be found in Table \ref{19-results}.

Our first set of results on the data set of 64 protein-nucleic acid complexes is shown in Table \ref{64-results}. We compare our findings to the flexibility-rigidity index (FRI) and multiscale FRI (mFRI) from Opron et al. \cite{opron2016} and the benchmark GNM \cite{yang2006}. The PSL model achieves a higher average Pearson correlation coefficient (PCC) than all other compared models for the M1, M2, and M3 representations. Most notably, our results using PSL features show an improvement over the GNM results by 21\% for the M3 model.

\begin{table}[H]
    \centering
    \begin{tabular}{c|c|c|c|c}
    \hline
    Model & PSL & mFRI \cite{opron2016} & FRI \cite{opron2016} & GNM \cite{yang2006} \\
    \hline
    M1  & \textbf{0.683} & 0.666 & 0.620 & 0.59 \\
    M2  & \textbf{0.669} & 0.668 & 0.612 & 0.58 \\
    M3  & \textbf{0.669} & 0.620 & 0.555 & 0.55 \\
    \hline
    \end{tabular}
\caption{Average Pearson correlation coefficients (PCC) on three coarse-grained models for a set of 64 protein-nucleic acid structures \cite{yang2006}. Results are shown for our PSL model compared to the flexibility-rigidity index (FRI) and multiscale FRI (mFRI) models by Opron et al. \cite{opron2016} and the Gaussian network model (GNM) \cite{yang2006}. The mFRI results shown were produced by a model using two kernels.}
\label{64-results}
\end{table}

In addition to the results on the entire data set, Opron et al. \cite{opron2016} also reported their average PCC for the subset of 19 nucleic-acid-only structures in the set of 64 complexes. Our results are compared to these in Table \ref{19-results}. Our PSL model achieved a higher average PCC value than that of the mFRI model for all three coarse-grained representations of the 19 structures in this set. In particular, we observe the biggest improvement for the M1 model, with our PSL model yielding an average PCC of 0.675, an 11\% increase from the mFRI result (PCC = 0.608).

\begin{table}[H]
    \centering
    \begin{tabular}{c|c|c}
    \hline
    Model & PSL & mFRI \cite{opron2016} \\
    \hline
    M1  & \textbf{0.675} & 0.608  \\
    M2  & \textbf{0.649} & 0.617  \\
    M3  & \textbf{0.629} & 0.603  \\
    \hline
    \end{tabular}
\caption{Average Pearson correlation coefficients (PCC) for three coarse-grained models of a set of 19 nucleic-acid-only structures \cite{yang2006} using our PSL model. Results are compared to those of a multiscale flexibility-rigidity index (mFRI) model using two kernels \cite{opron2016}.}
\label{19-results}
\end{table}

As shown in Table \ref{203-results}, our PSL model also achieved higher average PCC values on the set of 203 structures than the FRI and mFRI models \cite{opron2016}. Here, the most improvement is seen for the M3 representation, with the PSL model yielding an average PCC of 0.710, a 12\% increase compared to mFRI (PCC = 0.63).

\begin{table}[H]
    \centering
    \begin{tabular}{c|c|c|c}
    \hline
    Model & PSL & mFRI \cite{opron2016} & FRI \cite{opron2016} \\
    \hline
    M1  & \textbf{0.715} & 0.68 & 0.613 \\
    M2  & \textbf{0.718} & 0.67 & 0.625 \\
    M3  & \textbf{0.710} & 0.63 & 0.586 \\
    \hline
    \end{tabular}
    \caption{Average Pearson correlation coefficients (PCC) for three coarse-grained models of a set of 203 protein-nucleic acid structures \cite{opron2016} using our PSL model. Results are compared to those of a flexibility-rigidity index (FRI) model and a multiscale flexibility-rigidity-index (mFRI) model using two kernels  \cite{opron2016}.}
\label{203-results}
\end{table}

\subsection{Results on 126 Protein-RNA Structures}
\label{subsec:126-results}

% \textcolor{red}{More data sets? More computational results?} \\

In this section, we report our results for the prediction of B-factors of the 126 protein-RNA complexes described in Section \ref{subsec:data-sets}. Table \ref{tab:126-average-results} shows the average PCC values over all 126 structures for the three models. The PSL model performed fairly consistently for all three models of this data set, with the M3 model yielding the highest average PCC (PCC = 0.700) of the three representations. Again, we utilized the same PSL features for these predictions as in our previous experiments, as outlined in Section \ref{subsec:parameters}. Of course, these parameters can be further tuned to optimize the results on the 126 protein-RNA structures in this data set. Detailed results for the 126 protein-RNA structures are given in Table \ref{tab:126-all-results} in Appendix \ref{sec:appendix}. \\

\begin{table}[H]
    \centering
    \begin{tabular}{c|c}
    \hline
    Model & PSL \\
    \hline
    M1  & 0.669  \\
    M2  & 0.665  \\
    M3  & 0.700  \\
    \hline
    \end{tabular}
\caption{Average Pearson correlation coefficients (PCC) for three coarse-grained models of a set of 126 protein-RNA structures \cite{harini2025} using our PSL model.}
\label{tab:126-average-results}
\end{table}

\section{Discussion}\label{sec:discussion}

This section provides further discussion of the results presented in Section \ref{sec:results} and examines a few case studies of particular protein-nucleic acid complexes (and one nucleic-acid-only complex) to illustrate our model's strengths.

\subsection{Protein-Nucleic Acid Case Studies}
\label{subsec:case-studies}

% \textcolor{red}{Add more case studies here} \\

In addition to B-factor prediction results on the protein-nucleic acid data sets, in this section, we present case studies for selected protein-nucleic acid complexes to illustrate PSL model's advantage over mFRI on particular structures.

\begin{figure}[ht]
    \centering
    \includegraphics[width=0.9\textwidth]{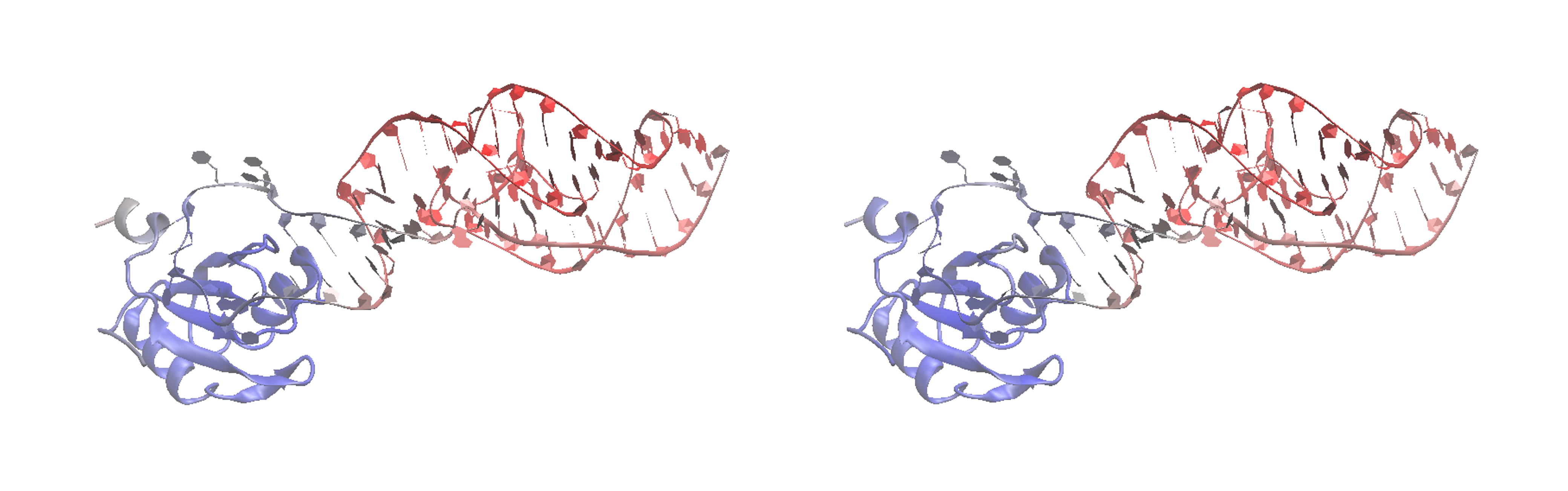}
    \includegraphics[width=0.9\textwidth]{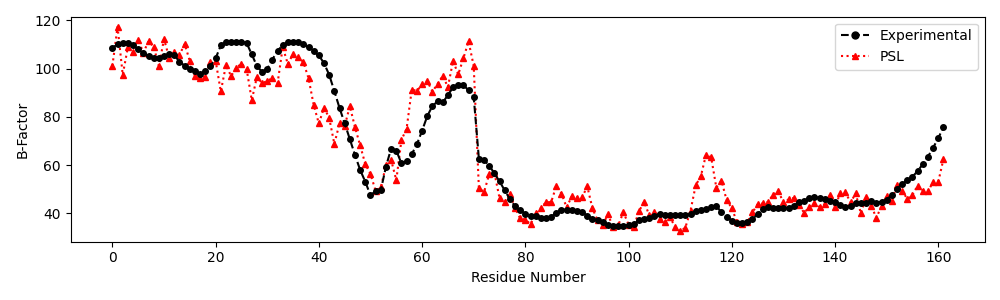}
    \caption{Top: 3D visualization of complex 1DRZ, with residues colored by experimental B-factors (left) and predicted B-factors using PSL (right). Bottom: Experimental and predicted B-factors for each residue of complex 1DRZ. The results shown use model M1 for this complex. Both chains A and B of the complex are depicted here, with chain A corresponding to the atoms numbered 1-71 above and chain B corresponding to the atoms numbered 72-162 above. Note that the numbering used here simply reflects the ordering of atoms in the M1 model as extracted from the PDB file---this is not a canonical numbering for the complex.}
    \label{fig:1drz-comparisons}
\end{figure}

The first case study concerns complex PDB: 1DRZ, a hepatitis $\delta$ virus ribozyme \cite{yang2006}. Ribozymes are RNA molecules (or protein-RNA complexes, such as complex 1DRZ) that can act as enzymes, catalyzing chemical reactions much like protein enzymes. However, the catalytic center of a ribozyme is made up solely of RNA, enabling catalysis without the need of a protein \cite{scott2007ribozymes}. Figure \ref{fig:1drz-comparisons} displays the predicted B-factors for this complex using the PSL model compared to the experimental B-factors. Additionally, we provide a 3D visualization of the ribozyme using the Visual Molecular Dynamics (VMD) software \cite{humphrey1996vmd}. Here, residues are colored by experimental or predicted B-factors. Less flexible regions are shown as blue (``colder" residues), and more flexible regions are shown as red (``warmer" residues). These predictions were generated using the M1 model for complex 1DRZ, which achieved the largest PCC value (PCC = 0.947) of all three coarse-grained models for this complex. The PSL model improves the result of the mFRI method (PCC = 0.846) \cite{opron2016} by 11\%. Both the mFRI and PSL models performed best with the M1 coarse-grained model for this complex. Given our large PCC on this complex, the visualizations reflect this accuracy, with the PSL model slightly over- and undershooting the B-factors for a few sections of this ribozyme. For instance, for the short alpha helix corresponding to the last numbered residues, the PSL model predicted lower B-factors, reflected by this structure's more saturated blue color in the 3D visualization.

\begin{figure}[h]
    \centering
    \includegraphics[width=0.8\textwidth]{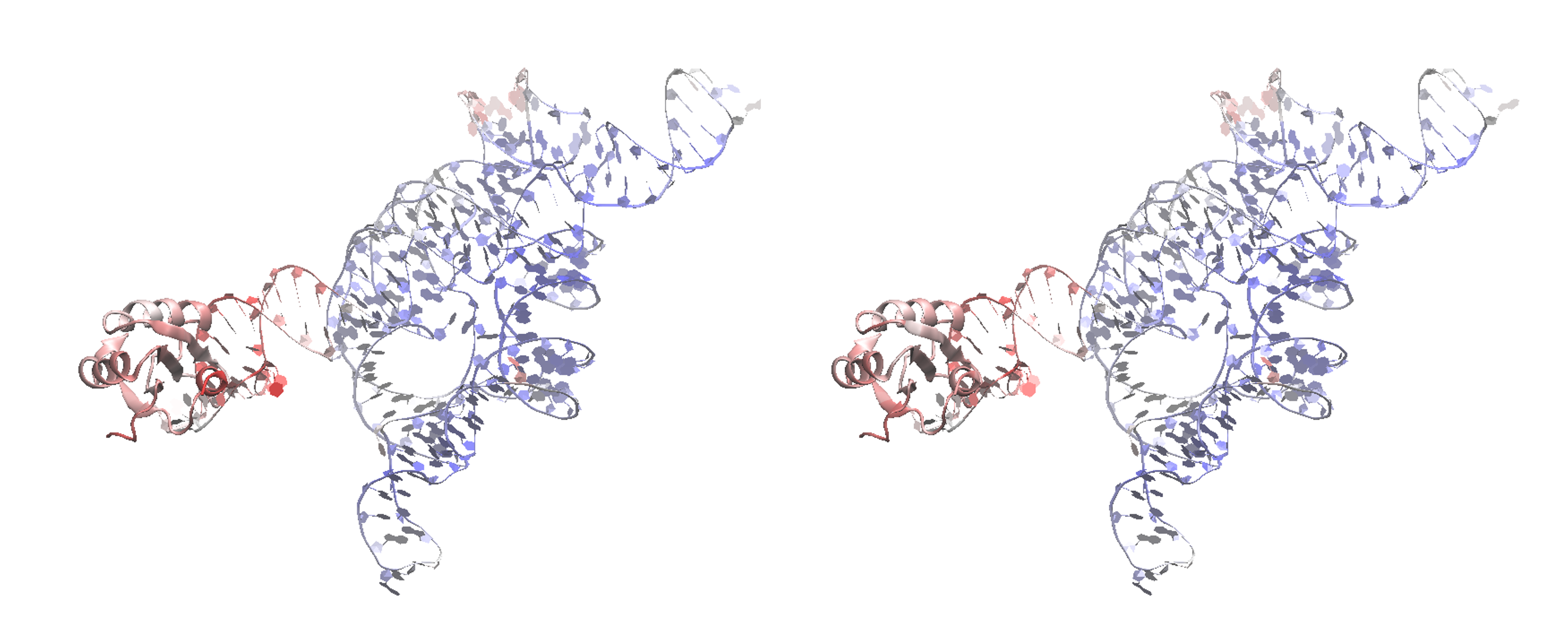}
    \includegraphics[width=0.9\textwidth]{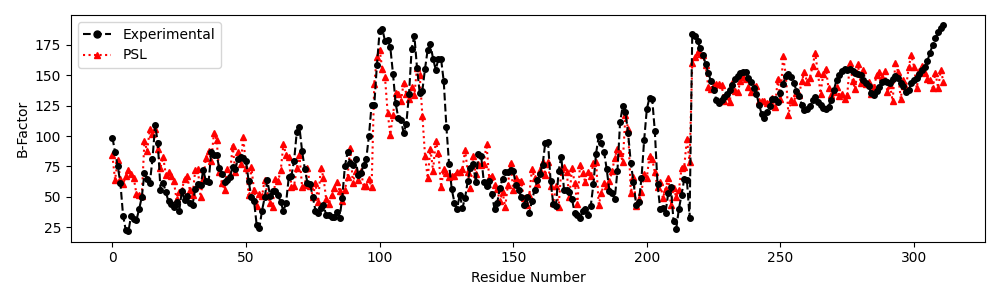}
    \caption{Top: 3D visualization of complex 1U6B, with residues colored by experimental B-factors (left) and predicted B-factors using PSL (right). Bottom: Experimental and predicted B-factors for each residue of complex 1U6B. The results shown use model M1 for this complex. Chains A, B, C, and D of the complex are depicted here, with chain A corresponding to the atoms numbered 218-312, chain B corresponding to the atoms numbered 0-195, and chain C corresponding to the atoms numbered 196-216. There is a single atom from Chain D numbered 217 in our model. Note that the numbering used here simply reflects the ordering of atoms in the M1 model as extracted from the PDB file---this is not a canonical numbering for the complex.}
    \label{fig:1u6b-comparisons}
\end{figure}

Another protein-nucleic acid complex for which the PSL model outperforms the mFRI method is PDB: 1U6B, a Group I intron \cite{yang2006}. Group I introns are ribozymes that perform RNA self-splicing---that is, they catalyze reactions to excise themselves from the precursor RNA \cite{cech1994representation, ohuchi2002modular}. Figure \ref{fig:1u6b-comparisons} shows the predicted and experimental B-factors for the complex 1U6B. The M1 model yielded the greatest PCC value of the three coarse-grained models on this complex as well, and the results shown are for this model. The PSL method achieved a PCC of 0.842, which is a 66\% improvement over the mFRI model's best result (PCC = 0.506) for this complex \cite{opron2016}. The mFRI model performed the best using the M3 model for this structure. The PSL model largely captures the variations in flexibility of complex 1U6B over all chains of the molecule, with some overestimations and underestimations at various areas, notably for one part of the flexible nucleic acid region corresponding roughly to residues 115-125 in Figure \ref{fig:1u6b-comparisons} (in chain B). The PSL model successfully predicts the increase in flexibility for the earlier residues of this flexible region but it underestimates the B-factors for the later residues of this region. Additionally, the B-factors for the flexible alpha helix corresponding to the last residues in Figure \ref{fig:1u6b-comparisons} (in chain A) are somewhat underestimated by the PSL model. Still, the model is able to perform well overall, particularly compared to mFRI.

\begin{figure}[h]
    \centering
    \includegraphics[width=0.9\textwidth]{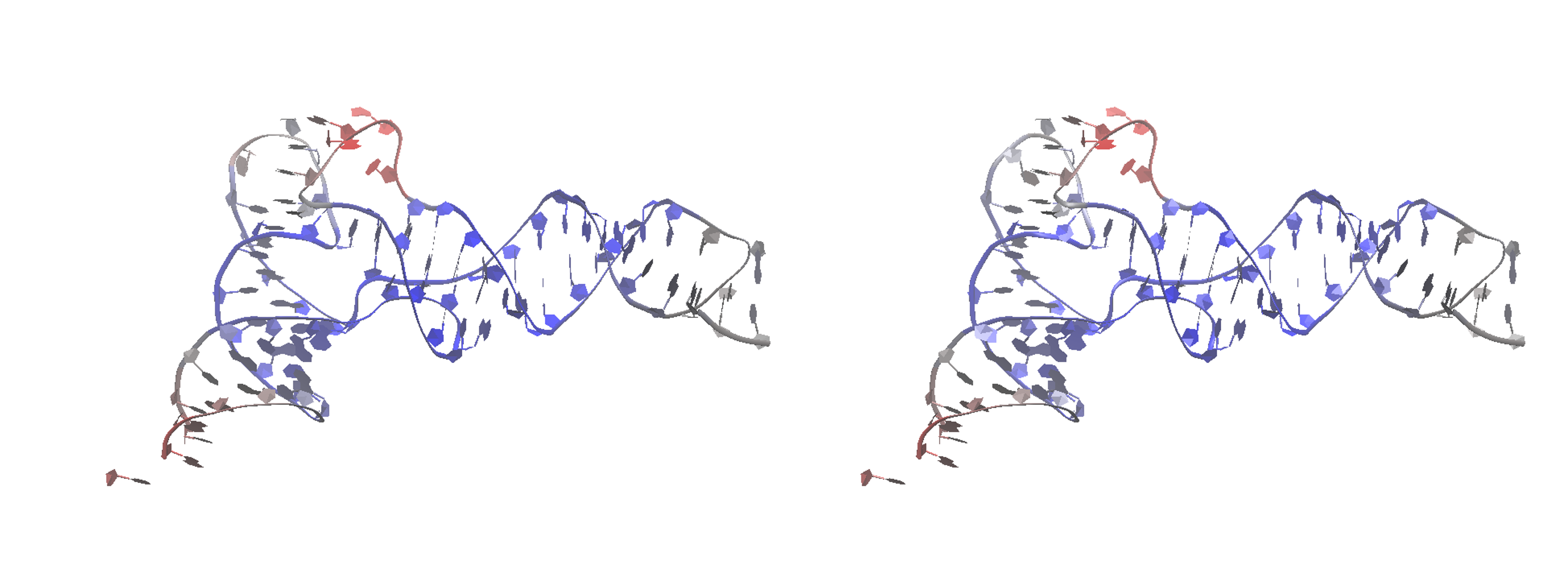}
    \includegraphics[width=0.9\textwidth]{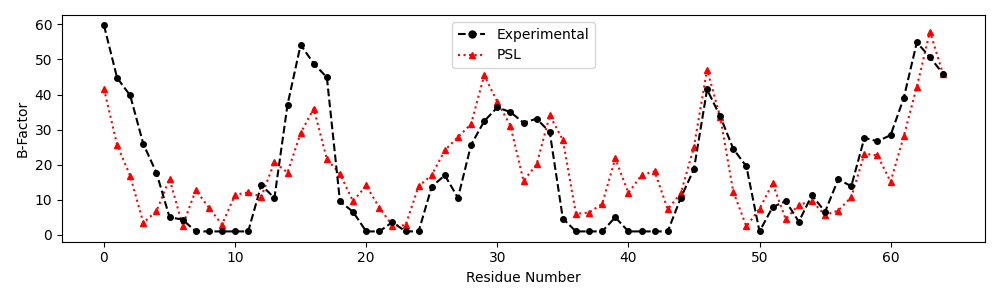}
    \caption{Experimental and predicted B-factors for each residue of nucleic-acid-only complex 2TRA. The results shown use model M1 for this complex, which consists of only one chain.}
    \label{fig:2tra-comparisons}
\end{figure}

Our third case study concerns the nucleic-acid-only complex PDB: 2TRA, an aspartic acid transfer RNA (tRNA) \cite{yang2006}. The primary function of tRNA lies in protein synthesis. Specifically, tRNA transports amino acids to the ribosome, where it then acts as a link between messenger RNA (mRNA) and the growing polypeptide chain. tRNA is also involved in other biological processes, including enzyme synthesis regulation, enzyme inhibition, and gene expression \cite{rich1976transfer, berg2021transfer}. Figure \ref{fig:2tra-comparisons} presents the predicted B-factors for complex 2TRA using PSL as well as experimental B-factors. The results displayed are those for the M1 model, which achieved the highest PCC of the three coarse-grained models for this complex. Our PSL method yielded a PCC value of 0.744, a 21\% improvement over mFRI (PCC = 0.614) \cite{opron2016}. The mFRI result from Opron et al. \cite{opron2016} is also for the M1 model, which resulted in the highest PCC value for mFRI for this structure, with the M2 model performing equally well. Overall, the PSL model predicts the changes in flexibility throughout the nucleic acid well, while missing some of the finer points. The model successfully predicts the B-factors for the flexible residue numbered 46 in Figure \ref{fig:2tra-comparisons} and also captures the flexible region corresponding to the last numbered residues of the molecule for this model, while underestimating residue 62. While the PSL model does predict the other trends in flexibility, it misses some of the extremes---it underestimates the B-factors for some flexible regions (residues 0-4, 14-17) and overestimates the B-factors for some rigid regions (residues 7-11, 35-43).

\subsection{Discussion of Results}
\label{subsec:discussion}

Our results in Section \ref{sec:results} demonstrate the improved performance of the PSL model over traditional GNM and FRI as well as mFRI, a multiscale method. The successes of the PSL and mFRI models with respect to the single-scale methods illustrate the benefit of multiscale models for flexibility analysis of protein-nucleic acids, where interactions take place across multiple length scales. We have considered multiple data sets of protein-nucleic acid complexes containing diverse structures, including some nucleic-acid-only complexes. Yang et al. \cite{yang2006} and Opron et al. \cite{opron2016} provide additional information about the complexes in the data sets of 64 and 203 structures.

In the set of 64 protein-nucleic acid complexes \cite{yang2006}, the coarse-grained molecule representations ranged from 61 atoms (for the M1 model representation for complex 1FIR) to 12378 atoms (for the M3 model representation for complexes 1S72, 1YHQ, and 1YIJ). Overall, the PSL model performed the best on average using the M1 model for this data set (average PCC = 0.683). In fact, the M1 model yielded the highest PCC value for all three complexes considered in our case studies in Section \ref{subsec:case-studies}. These three complexes, 1DRZ, 1U6B, and 2TRA, were represented by models with a relatively small number of atoms: 162, 312, and 65, respectively, for the M1 model. These model sizes reflect the sizes of the complexes themselves---complex 1DRZ has 95 amino acid residues and 72 nucleotides, complex 1U6B has 95 residues and 222 nucleotides, and complex 2TRA has 74 nucleotides \cite{yang2006}. Thus, it is perhaps unsurprising that the PSL model using the M1 representation (with the fewest number of atoms per nucleotide) was sufficient to capture the topological and geometric information of these structures.

Generally, the PSL model performed better for smaller structures (i.e., with fewer residues and/or nucleotides) than for larger structures for the 64-complex data set. This may be due to increased complexity of the larger molecules as well as the fact that we used the same filtration parameters for all complexes in the present work. Although our parameters may have generated features that encoded enough information about smaller molecules, adjusting the filtration radii to include greater length scales may improve the predictive performance of the PSL model on larger complexes by capturing potential longer-range interactions and additional structure around each atom. Additionally, although the M3 model had a lower PCC value on average for this data set, some of the larger molecules had better results with the M3 model than the M1 and M2 models. For instance, complex 1YIJ (a 50S ribosomal subunit with 3775 amino acid residues and 2876 nucleotides \cite{yang2006}) had a PCC of 0.498 using the M1 model (with 6636 atoms) and a PCC of 0.546 using the M2 model (with 9507 atoms), but using the M3 model (with 12378 atoms) increased its PCC to 0.668. Consequently, we suggest that strategically choosing an appropriate coarse-grained model and tuning PSL filtration parameters based on molecular structure can improve flexibility prediction accuracy on a given complex.

The set of 203 structures \cite{opron2016} had many smaller molecules, with the complex 1VTG containing only seven atoms in its M1 representation. This data set also lacked such large molecules as the 64-structure data set---here, the complex with the most atoms in its coarse-grained representation was 4DQQ, with 1278 atoms using the M3 model. Given the above discussion, it is likely not surprising that the PSL model thus performed better on average for the 203-complex data set. This time, the M2 model yielded the best result (which was: average PCC = 0.718), although the M1 and M3 models showed very similar performance (which was: M1 average PCC = 0.715, M3 average PCC = 0.710). We do observe some similar trends for this data set as for the set of 64 complexes, with the PSL model generally achieving better predictive performance on the smaller structures.

In addition to the data sets of 64 and 203 protein-nucleic acid complexes, which have existing results in the literature for GNM, FRI, and mFRI, we also predicted B-factors for a data set of 126 protein-RNA complexes from a set of 710 mutations on these structures \cite{liu2025, harini2025}. Detailed results and descriptions of the coarse-grained models for these 126 complexes are given in Appendix \ref{sec:appendix}. This data set did not contain structures as small as those in the set of 203 complexes nor structures as large as those in the set of 64 complexes. For this data set, the PSL model performed best on average using the M3 representation (average PCC = 0.700). Furthermore, while the structures on which PSL performed the best were similar in size to the most successful complexes in the other two data sets, many of the larger structures in this data set had better results compared to similarly sized complexes in the other data sets.

Overall, the results of our PSL model on these three data sets demonstrate its applicability for flexibility analysis of diverse protein-nucleic acid complexes, as evidenced by its increased predictive accuracy over traditional GNM and mFRI, a state-of-the-art multiscale model. Our case studies further illustrate the advantages of PSL for particular significant structures. Additionally, we propose that the performance of the PSL model on larger complexes may be improved by parameter tuning and strategic coarse-grained model selection.

\section{Methods}\label{sec:methods}

\subsection{Persistent Sheaf Laplacians}\label{subsec:psl}

Many of the techniques used in the present work follow those in our prior paper \cite{hayes2025}, with some adjustments made for the more complex nucleic acid structures. The algorithms for data preprocessing are described in Section \ref{subsec:data-sets}, with particular discussion of three coarse-grained models from Yang et al. \cite{yang2006} that we adopted for our analyses. The specific persistent sheaf Laplacian features are also detailed in this paper in Section \ref{subsec:parameters}. In this section, we will briefly review the persistent sheaf Laplacian theory and provide more detail about the feature generation. The full details and comprehensive review of the persistent sheaf Laplacian theory and algorithms can be found in the original paper by Wei and Wei \cite{wei2025persistent} and our earlier paper \cite{hayes2025}. Other graph-based approaches incorporating Laplacians include the work\cite{hayes2023}.

Persistent sheaf Laplacians \cite{wei2025persistent} (PSLs) extend the theory of persistent Laplacians on simplicial complexes and simplicial chain complexes to cellular sheaves and sheaf cochain complexes. The motivation for the extension to sheaves is, in part, their ability to embed additional non-spatial information in a simplicial complex. In molecular science, this information may be derived from molecular structures; one example of such information is an atom's partial charge. In contrast to many other persistent topological Laplacians, which can describe a molecule as a whole, persistent sheaf Laplacians provide information about local (i.e., atom-specific) topology and geometry in a molecule.

To formalize the above, we will first introduce persistent homology, which is defined with respect to simplicial complexes. Given a finite set $V$ (for example, a set of atoms in a molecule), we can define a simplicial complex $X$ on $V$ as a collection of subsets of $V$ satisfying the following criterion: if a set $\sigma$ is in $X$, then any subset of $\sigma$ must also be in $X$. The sets $\sigma$ are called simplices, and a simplex with $q+1$ elements is called a $q$-simplex. Furthermore, if a simplex $\sigma \subset \tau$, then $\sigma$ is called a face of $\tau$, and this face relation is defined by $\sigma \leqslant \tau$. We can define subcomplexes as follows: if $X$ and $Y$ are both simplicial complexes with $X \subset Y$, then $X$ is considered a subcomplex of $Y$.

Given a simplicial complex $X$, one may define the simplicial chain complex associated with $X$, which is denoted as
\[
    \begin{tikzcd}[column sep = large]
        \centering
    \cdots \arrow[r, "\partial_{3}"] & 
    C_{2}(X) \arrow[r, "\partial_{2}"] & 
    C_{1}(X)  \arrow[r, "\partial_1"] & 
    C_{0}(X) \arrow[r] & 
    0.
    \end{tikzcd} 
\]

$C_q(X)$ is a real vector space generated by $q$-simplices of $X$, and an element $\sigma \in C_q(X)$ is called a $q$-chain. The operator $\partial_q: C_q(X) \to C_{q-1}(X)$ is a linear map, called a boundary map, defined as follows:
\begin{align*}
    \partial_q [v_{a_0}, \dots, v_{a_q}] = \sum_i (-1)^i[v_{a_0}, \dots, \hat{v}_{a_i}, \dots, v_{a_q}].
\end{align*}

Here, the notation $ \hat{v}_{a_i}$ means that the element $v_{a_i}$ is deleted. Because our finite set $V$ is totally ordered, the boundary operator $\partial_q$ is well defined. We can then define the $q$-th homology group of the chain complex by $H_q=\ker \partial_{q}/\text{im} \partial_{q+1}$. We have $\partial^2 = \partial_{q} \circ \partial_{q+1} = 0$, so $H_q$ is also well defined.

Now, if we let $X$ be a subcomplex of $Y$, we also have inclusions of the chain groups, with the inclusion map denoted by $\iota: C_q(X) \hookrightarrow C_q(Y)$. Thus, we can construct the following diagram: 

\[
    \begin{tikzcd}
        \cdots \arrow[r, "\partial_{q+2}^X"]
        & C_{q+1}(X) \arrow[r, "\partial_{q+1}^X"] \arrow[d, hook, dashed] 
          & C_{q}(X) \arrow[r, "\partial_q^X"] \arrow[d, hook, dashed] 
            & C_{q-1}(X) \arrow[r, "\partial_{q-1}^X"] \arrow[d, dashed, hook]
              & \cdots
            \\
        \cdots \arrow[r, "\partial_{q+2}^Y"]
        & C_{q+1}(Y) \arrow[r, "\partial_{q+1}^Y"] 
          & C_q(Y) \arrow[r, "\partial_q^Y"] 
            & C_{q-1}(Y) \arrow[r, "\partial_{q-1}^Y"]
              & \cdots 
    \end{tikzcd}
\]

Then, we have a map $\iota^{\bullet}$ induced by the inclusion $\iota$, defined as $\iota^{\bullet}: H_q(X) \to H_q(Y)$. The image
\begin{align*}
    \iota^{\bullet} (H_q(X))
\end{align*}
 is the $q$-th persistent homology group for the simplicial complex pair $(X,Y)$. The ranks of these persistent homology groups are called ($q$-th) persistent Betti numbers $\beta_{q}^{X,Y}$ \cite{wei2025persistent}, which are used in topological data analysis to track topological features that persist in the filtration (i.e., at multiple scales). If the input point cloud is given by a set of atoms in a molecule, the Betti numbers can capture multiscale information about the molecule's structure and interactions.

Persistent Laplacians \cite{wang2020persistent}---in particular, their non-zero eigenvalues---can capture even more information from an input point cloud. Given a persistent homology group $\iota^{\bullet}(H_q(X))$ as defined above, a persistent Laplacian is a positive semi-definite operator with a kernel isomorphic to that group. To construct the persistent Laplacian, first define $C_{q+1}^{X,Y}=\{c \in C_{q+1}(Y) \mid \partial_{q+1}^Y(c) \in C_q(X) \}$, and let $\partial_{q+1}^{X,Y}$ be the restriction of $\partial^Y_{q+1}$ to $C_{q+1}^{X,Y}$. Note that each $C_q(X)$ is generated by $q$-simplices, equipping it with a canonical inner product. Now, we can define the $q$-th persistent Laplacian $\Delta_{q}^{X,Y}$ as
\begin{align*}
    \Delta_{q}^{X,Y} = \partial_{q+1}^{X,Y} (\partial_{q+1}^{X,Y})^{\dag} + (\partial^X_{q})^{\dag} \partial^X_{q},
\end{align*}
where $\dag$ is the adjoint of a linear morphism. Many variants of persistent Laplacian methods exist, and examining the eigenvalues of these various persistent Laplacians enables machine learning algorithms to learn multiscale information about point clouds in addition to that available from persistent homology.

Now, while persistent Laplacians as defined above are built on simplicial chain complexes, persistent sheaf Laplacian theory \cite{wei2025persistent} generalizes this idea to cellular sheaves and sheaf cochain complexes. A cellular sheaf $\mathscr{F}$ can be conceptualized as a simplicial complex $X$, where we assign (i) a finite-dimensional vector space $\mathscr{S}(\sigma)$ to each simplex $\sigma$ of $X$ and (ii) a linear morphism $\mathscr{S}_{\sigma \leqslant \tau}$ of vector spaces to each face relation $\sigma \leqslant \tau$. The vector space $\mathscr{S}(\sigma)$ is called the stalk of $\mathscr{S}$ over $\sigma$, and the morphism $\mathscr{S}_{\sigma \leqslant \tau}$ is called the restriction map of the face relation $\sigma \leqslant \tau$. Furthermore, the restriction maps behave nicely across face relations; that is, $\mathscr{S}(\sigma)$ satisfies the rule that if $\rho \leqslant \sigma \leqslant \tau$, then $\mathscr{S}_{\rho \leqslant \tau} = \mathscr{S}_{\sigma \leqslant \tau} \mathscr{S}_{\rho \leqslant \sigma}$, where $\mathscr{S}_{\sigma \leqslant \sigma}$ refers to the identity map of $\mathscr{S}(\sigma)$.

Intuitively, stalks serve as information about each simplex; in the case where the input point cloud represents atoms in a molecule, a stalk may capture non-spatial atomic information, such as partial charges, atomic weights, etc. We can view restriction maps as describing how the additional information stored by stalks interacts across simplexes.

Now, we may construct a sheaf cochain complex
\[
    \begin{tikzcd}
        \centering
    0 \arrow[r] & 
    C^0(X; \mathscr{S}) \arrow[r, "d"] & 
    C^1(X; \mathscr{S}) \arrow[r, "d"] & 
    C^2(X; \mathscr{S}) \arrow[r, "d"] & 
    \cdots,
    \end{tikzcd} 
\]
where each $q$-th sheaf cochain group $C^q(X; \mathscr{S})$ is defined as the direct sum of stalks over all $q$-dimensional simplices. The coboundary maps $d$ require obtaining a signed incidence relation by globally orienting the simplicial complex $X$, where the relation assigns an integer $[\sigma: \tau]$ to each face relation $\sigma \leqslant \tau$. Thus, we can define the coboundary map $d^{q}: C^q(X;\mathscr{S}) \to C^{q+1}(X;\mathscr{S})$ as
\begin{align*}
    d^q \vert_{\mathscr{S}(\sigma)} = \sum_{\sigma \leqslant \tau} [\sigma: \tau] \mathscr{S}_{\sigma \leqslant \tau}.
\end{align*}

To define the persistent sheaf Laplacian, we must again consider subcomplexes $X \subset Y$, now equipped with sheaf $\mathscr{F}$ on $X$ and sheaf $\mathscr{G}$ on $Y$. Here, we let the sheaves be defined such that the stalks and restriction maps of $X$ and $Y$  are identical. Thus, we have a sheaf cochain complex for $X$ and one for $Y$, with inclusion maps between their corresponding cochain groups. Further assuming that each stalk is an inner product space, we can define $\Theta_{\mathscr{F}, \mathscr{G}}^{q+1} = \{x \in C^{q+1}(Y; \mathscr{G}) \mid (d^{q}_{\mathscr{G}})^{\dag}(x) \in C^{q}(X; \mathscr{F})\}$, where $d^{q}_{\mathscr{F}, \mathscr{G}}$ is the adjoint of $\pi (d^{q}_{\mathscr{G}})^{\dag}\vert_{\Theta^{q+1}_{\mathscr{F}, \mathscr{G}}}: \Theta^{q+1}_{\mathscr{F}, \mathscr{G}} \to C^q(X;\mathscr{F})$ and $\pi$ is the projection map from $C^q(Y; \mathscr{G})$ to $C^q(X; \mathscr{F})$ (its subspace). The $q$-th persistent sheaf Laplacian $\Delta_q^{\mathscr{F}, \mathscr{G}}$ is then defined as
\begin{align*}
    \Delta_q^{\mathscr{F}, \mathscr{G}} = (d^{q}_{\mathscr{F}, \mathscr{G}})^{\dag} d^{q}_{\mathscr{F}, \mathscr{G}} +  d^{q-1}_{\mathscr{F}}(d^{q-1}_{\mathscr{F}})^{\dag}.
\end{align*}
Note that when $\mathscr{F}= \mathscr{G}$, the persistent sheaf Laplacian becomes the sheaf Laplacian of $\mathscr{F}$. Also, when $\mathscr{F}$ and $\mathscr{G}$ are constant, the persistent sheaf Laplacian is equal to the persistent Laplacian $\Delta_{q}^{X,Y}$ defined above.

\subsection{Model Construction and Feature Generation}\label{subsec:psl-details}

Given the theoretical foundation of PSLs in the previous section, in this section we will outline the specifics of the PSL model used in this paper, as well as how we generated machine learning features using this model.

In general, persistent homology examines how topological features evolve across a nested sequence of subcomplexes, called a filtration. Given a point cloud, a common method for constructing a filtration is by varying a radius parameter from each point and including additional simplices once the radius parameter exceeds the distance from our point to each of those simplices. In the present work, the filtration parameter is the distance (in \r{A}) from each atom. At each radius, one may compute topological invariants of the corresponding subcomplex. Then, how these invariants change over the filtration can be analyzed. In a molecule, this can allow the examination of interactions and structures at multiple length scales. However, many of the typical topological tools provide global information, whereas protein-nucleic acid flexibility analysis requires the knowledge of atom-specific information to predict individual B-factors. Persistent sheaf Laplacians provide the necessary localization for this analysis. To construct a filtration of sheaves, as used in persistent sheaf cohomology \cite{russold2022persistent} and persistent sheaf Laplacians, we begin by constructing a filtration of simplicial complexes as above, but additionally construct a sheaf for each complex. Then, by computing the eigenvalues of persistent sheaf Laplacians, we can obtain topological invariants (from the harmonic spectra) and geometric information (from the non-harmonic spectra) of the data \cite{wei2025persistent2}.

Specifically, for the present work, we begin with a point cloud consisting of the atoms in a particular coarse-grained model (M1, M2, or M3) of a protein-nucleic acid complex. With the goal of predicting the B-factor for each atom, we construct features for each atom using PSLs. For a given atom $A$ in the point cloud, we first designate a cutoff distance for the point cloud so that we only consider other atoms within that distance from $A$ when we construct our complex. Then, we determine the set of radii that will generate our filtration---for this work, we used 6\r{A}, 12\r{A}, and 18\r{A}. For each radius, we construct an alpha complex $X$ (again, only considering atoms within the prescribed cutoff distance from $A$). To construct a cellular sheaf on $X$, we assign a label $q_i$ to each atom $v_i$ in $X$, and then let each stalk be $\mathbb{R}$. To define the restriction maps, let $r_{ij}$ be the length of simplex $v_iv_j$ (i.e., the 1-simplex between atoms $v_i$ and $v_j$). For face relations of the form $v_i \leqslant v_iv_j$, we define the restriction map as scalar multiplication by $q_j/r_{ij}$. For face relations of the form $v_iv_j \leqslant v_iv_jv_k$, we define the restriction map as scalar multiplication by $q_k/(r_{ik}r_{jk})$. (Further motivation behind the definition of restriction maps in this way can be found in our previous paper \cite{hayes2025} and the introduction by Wei et al. \cite{wei2025persistent}.) In order to distinguish atom $A$ from the other model atoms in $X$, the label $q_i$ of $A$ is set to 0, and the labels of all other model atoms in $X$ are set to 1. By constructing sheaf Laplacians for atom $A$ for a given radius, and then computing the eigenvalues of the sheaf Laplacians, we can generate features for $A$ using these eigenvalues (the particular set of features used in this work is detailed in Section \ref{subsec:parameters}). Then, by varying the radius and computing sheaf Laplacians across the filtration, we obtain additional features for each radius. We can repeat this process for all atoms in the given coarse-grained model to enable the B-factor prediction of each atom.

\section{Conclusion}
 Persistent topological Laplacians are a new generation of multiscale spectral algorithms \cite{su2025topological}.  Unlike other topological techniques, the persistent sheaf Laplacian (PSL) offers atom-specific localized information of a biomolecular complex\cite{wei2025persistent}. The atom-specific and multiscale nature of persistent sheaf Laplacian features enables detailed atomistic analysis, which is crucial for understanding multiscale molecular flexibility and long-term conformational dynamics. Both molecular flexibility and the long-term conformational dynamics are very important for their biological functions and are the subjects of biophysical studies in the past decades.  
 
 In this study, we introduce a PSL model to predict the B-factors of protein–nucleic acid complexes, a task that presents unique challenges due to their multiscale nature and structural diversity. Our results across multiple benchmark datasets demonstrate that PSL not only surpasses traditional models like GNM and mFRI in predictive accuracy but also maintains strong performance on nucleic-acid-only structures. 
 Furthermore, we have given a few case studies to further illustrate the model’s effectiveness in capturing biologically relevant flexibility patterns. The promising performance of our PSL model suggests its potential for further applications in structural biology, including the prediction of mutation effects, binding affinity changes, and the design of biomolecular therapeutics.

\section*{Acknowledgments}

This work is supported in part by NSF grants DMS-2052983 and DMS-2512644, NIH grant R35GM148196, and Michigan State University Research Foundation. 

\section*{Data and Code Availability}

The code for the PSL model, including an example of PSL feature generation and B-factor prediction, is available at\\
\href{https://github.com/weixiaoqimath/persistent_sheaf_Laplacians}{https://github.com/weixiaoqimath/persistent\_sheaf\_Laplacians}.

\section*{Author Contributions}

Nicole Hayes was responsible for the software, computational experiments and writing. Ekaterina Merkurjev and Guo-Wei Wei were responsible for methodology, reviewing, editing, some writing and supervision.

%Conception and design: Guo-Wei Wei. Sample preparation and collection of data: Nicole Hayes. Algorithm implementation: Xiaoqi Wei,Hongsong Feng. Analysis and interpretation of data: Nicole Hayes, Guo-Wei Wei. Supervision: Ekaterina Merkurjev, Guo-Wei Wei. Manuscript preparation: Nicole Hayes, Xiaoqi Wei, Hongsong Feng, Ekaterina Merkurjev, Guo-Wei Wei. All authors contributed to the article and approved the submitted version.

\section*{Conflict of Interest}
 The authors have no conflicts to disclose. \\

\bibliography{bib_draft.bib}

\begin{thebibliography}{10}

\bibitem{petsko2004protein}
Gregory~A. Petsko and Dagmar Ringe.
\newblock {\em Protein structure and function}.
\newblock New Science Press, 2004.

\bibitem{ollis1987structural}
David~L. Ollis and Stephen~W. White.
\newblock Structural basis of protein-nucleic acid interactions.
\newblock {\em Chemical Reviews}, 87(5):981--995, 1987.

\bibitem{opron2016}
Kristopher Opron, Kelin Xia, Zach Burton, and Guo-Wei Wei.
\newblock Flexibility–rigidity index for protein–nucleic acid flexibility and fluctuation analysis.
\newblock {\em Journal of Computational Chemistry}, 37(14):1283--1295, 2016.

\bibitem{Anfinsen1973}
Christian~B. Anfinsen.
\newblock Principles that govern the folding of protein chains.
\newblock {\em Science}, 181(4096):223--230, 1973.

\bibitem{radivojac2004protein}
Predrag Radivojac, Zoran Obradovic, David~K. Smith, Guang Zhu, Slobodan Vucetic, Celeste~J. Brown, J.~David Lawson, and A.~Keith Dunker.
\newblock Protein flexibility and intrinsic disorder.
\newblock {\em Protein Science}, 13(1):71--80, 2004.

\bibitem{branden2012introduction}
Carl~Ivar Branden and John Tooze.
\newblock {\em Introduction to protein structure}.
\newblock Garland Science, 2012.

\bibitem{Frauenfelder1991}
Hans Frauenfelder, Stephen~G. Sligar, and Peter~G. Wolynes.
\newblock The energy landscapes and motions of proteins.
\newblock {\em Science}, 254(5038):1598--1603, 1991.

\bibitem{hyeon2006size}
Changbong Hyeon, Ruxandra~I. Dima, and D.~Thirumalai.
\newblock Size, shape, and flexibility of {RNA} structures.
\newblock {\em The Journal of Chemical Physics}, 125(19):194905, 11 2006.

\bibitem{ma2005usefulness}
Jianpeng Ma.
\newblock Usefulness and limitations of normal mode analysis in modeling dynamics of biomolecular complexes.
\newblock {\em Structure}, 13(3):373--380, 2005.

\bibitem{vihinen1994accuracy}
Mauno Vihinen, Esa Torkkila, and Pentti Riikonen.
\newblock Accuracy of protein flexibility predictions.
\newblock {\em Proteins: Structure, Function, and Bioinformatics}, 19(2):141--149, 1994.

\bibitem{jacobs2001protein}
Donald~J. Jacobs, Andrew~J. Rader, Leslie~A. Kuhn, and Michael~F. Thorpe.
\newblock Protein flexibility predictions using graph theory.
\newblock {\em Proteins: Structure, Function, and Bioinformatics}, 44(2):150--165, 2001.

\bibitem{camps2009flexserv}
Jordi Camps, Oliver Carrillo, Agust{\'\i} Emperador, Laura Orellana, Adam Hospital, Manuel Rueda, Damjan Cicin-Sain, Marco D'Abramo, Josep~Llu{\'\i}s Gelp{\'\i}, and Modesto Orozco.
\newblock Flex{S}erv: an integrated tool for the analysis of protein flexibility.
\newblock {\em Bioinformatics}, 25(13):1709--1710, 2009.

\bibitem{McCammon1977}
J.~Andrew McCammon, Bruce~R. Gelin, and Martin Karplus.
\newblock Dynamics of folded proteins.
\newblock {\em Nature}, 267(5612):585--590, 1977.

\bibitem{huber1983}
Robert Huber and William~S. Bennett~Jr.
\newblock Functional significance of flexibility in proteins.
\newblock {\em Biopolymers}, 22(1):261--279, 1983.

\bibitem{dudko2006intrinsic}
Olga~K. Dudko, Gerhard Hummer, and Attila Szabo.
\newblock Intrinsic rates and activation free energies from single-molecule pulling experiments.
\newblock {\em Physical Review Letters}, 96:108101, Mar 2006.

\bibitem{sun2019}
Zhoutong Sun, Qian Liu, Ge~Qu, Yan Feng, and Manfred~T. Reetz.
\newblock Utility of {B}-factors in protein science: Interpreting rigidity, flexibility, and internal motion and engineering thermostability.
\newblock {\em Chemical Reviews}, 119(3):1626--1665, 2019.
\newblock PMID: 30698416.

\bibitem{yuan2005prediction}
Zheng Yuan, Timothy~L. Bailey, and Rohan~D. Teasdale.
\newblock Prediction of protein {B}-factor profiles.
\newblock {\em Proteins: Structure, Function, and Bioinformatics}, 58(4):905--912, 2005.

\bibitem{hayes2025}
Nicole Hayes, Xiaoqi Wei, Hongsong Feng, Ekaterina Merkurjev, and Guo-Wei Wei.
\newblock Persistent sheaf {Laplacian} analysis of protein flexibility.
\newblock {\em The Journal of Physical Chemistry B}, 129(17):4169--4178, 2025.

\bibitem{Atilgan2001}
Ali~Rana Atilgan, Stewart Durell, Robert Jernigan, Melik Demirel, Özlem Keskin, and Ivet Bahar.
\newblock Anisotropy of fluctuation dynamics of proteins with an elastic network model.
\newblock {\em Biophysical Journal}, 80(1):505--515, 2001.

\bibitem{flory1976statistical}
Paul~J Flory.
\newblock Statistical thermodynamics of random networks.
\newblock {\em Proceedings of the Royal Society of London. A. Mathematical and Physical Sciences}, 351(1666):351--380, 1976.

\bibitem{Bahar1997}
Ivet Bahar, Ali~Rana Atilgan, and Burak Erman.
\newblock Direct evaluation of thermal fluctuations in proteins using a single-parameter harmonic potential.
\newblock {\em Folding and Design}, 2(3):173--181, 1997.

\bibitem{haliloglu1997}
Turkan Haliloglu, Ivet Bahar, and Burak Erman.
\newblock Gaussian dynamics of folded proteins.
\newblock {\em Physical Review Letters}, 79:3090--3093, Oct 1997.

\bibitem{xia2013multiscale}
Kelin Xia, Kristopher Opron, and Guo-Wei Wei.
\newblock Multiscale multiphysics and multidomain models—flexibility and rigidity.
\newblock {\em The Journal of Chemical Physics}, 139(19), 2013.

\bibitem{opron2014fast}
Kristopher Opron, Kelin Xia, and Guo-Wei Wei.
\newblock Fast and anisotropic flexibility-rigidity index for protein flexibility and fluctuation analysis.
\newblock {\em The Journal of Chemical Physics}, 140(23):234105, 2014.

\bibitem{xia2015multiscale}
Kelin Xia, Kristopher Opron, and Guo-Wei Wei.
\newblock Multiscale {Gaussian} network model ({mGNM}) and multiscale anisotropic network model ({mANM}).
\newblock {\em The Journal of Chemical Physics}, 143(20):204106, 11 2015.

\bibitem{Bahar1998}
Ivet Bahar, Ali~Rana Atilgan, Melik~C. Demirel, and Burak Erman.
\newblock Vibrational dynamics of folded proteins: Significance of slow and fast motions in relation to function and stability.
\newblock {\em Physical Review Letters}, 80:2733--2736, Mar 1998.

\bibitem{Hinsen1998}
Konrad Hinsen.
\newblock Analysis of domain motions by approximate normal mode calculations.
\newblock {\em Proteins: Structure, Function, and Bioinformatics}, 33(3):417--429, 1998.

\bibitem{Li2002}
Guohui Li and Qiang Cui.
\newblock A coarse-grained normal mode approach for macromolecules: An efficient implementation and application to {Ca$^{2+}$-ATPase}.
\newblock {\em Biophysical Journal}, 83(5):2457--2474, 2024/07/24 2002.

\bibitem{Tama2001}
Florence Tama and Yves-Henri Sanejouand.
\newblock {Conformational change of proteins arising from normal mode calculations}.
\newblock {\em Protein Engineering, Design and Selection}, 14(1):1--6, 01 2001.

\bibitem{Yang2008Coarse}
Lee-Wei Yang and Choon-Peng Chng.
\newblock Coarse-grained models reveal functional dynamics - {I}. {E}lastic network models – theories, comparisons and perspectives.
\newblock {\em Bioinformatics and Biology Insights}, 2:BBI.S460, 2008.
\newblock PMID: 19812764.

\bibitem{park2013coarse}
Jun-Koo Park, Robert Jernigan, and Zhijun Wu.
\newblock Coarse grained normal mode analysis vs. refined {Gaussian} network model for protein residue-level structural fluctuations.
\newblock {\em Bulletin of Mathematical Biology}, 75:124--160, 2013.

\bibitem{hinsen2008structural}
Konrad Hinsen.
\newblock Structural flexibility in proteins: impact of the crystal environment.
\newblock {\em Bioinformatics}, 24(4):521--528, 2008.

\bibitem{kondrashov2007protein}
Dmitry~A. Kondrashov, Adam~W. Van~Wynsberghe, Ryan~M. Bannen, Qiang Cui, and George~N. Phillips.
\newblock Protein structural variation in computational models and crystallographic data.
\newblock {\em Structure}, 15(2):169--177, 2007.

\bibitem{feng2025multiscale}
Hongsong Feng, Jeffrey~Y Zhao, and Guo-Wei Wei.
\newblock Multiscale differential geometry learning for protein flexibility analysis.
\newblock {\em Journal of Computational Chemistry}, 46(7):e70073, 2025.

\bibitem{opron2015}
Kristopher Opron, Kelin Xia, and Guo-Wei Wei.
\newblock Communication: Capturing protein multiscale thermal fluctuations.
\newblock {\em The Journal of Chemical Physics}, 142(21):211101, 06 2015.

\bibitem{yang2006}
Lee-Wei Yang, A.~J. Rader, Xiong Liu, Cristopher~Jon Jursa, Shann~Ching Chen, Hassan~A. Karimi, and Ivet Bahar.
\newblock {\textit{o}}{GNM}: online computation of structural dynamics using the {G}aussian network model.
\newblock {\em Nucleic Acids Research}, 34(suppl\_2):W24--W31, 07 2006.

\bibitem{peng2021flexibility}
Jiajie Peng, Jinjin Yang, D.~Vijay Anand, Xuequn Shang, and Kelin Xia.
\newblock Flexibility and rigidity index for chromosome packing, flexibility and dynamics analysis.
\newblock {\em Frontiers of Computer Science}, 16(4):164902, 2021.

\bibitem{carlsson2009topology}
Gunnar Carlsson.
\newblock Topology and data.
\newblock {\em Bulletin of the American Mathematical Society}, 46(2):255--308, 2009.

\bibitem{edelsbrunner2008persistent}
Herbert Edelsbrunner and John Harer.
\newblock Persistent homology--a survey.
\newblock {\em Contemporary Mathematics}, 453(26):257--282, 2008.

\bibitem{wee2025review}
JunJie Wee and Jian Jiang.
\newblock A review of topological data analysis and topological deep learning in molecular sciences.
\newblock {\em Journal of Chemical Information and Modeling}, accepted, 2025.

\bibitem{wang2020persistent}
Rui Wang, Duc~Duy Nguyen, and Guo-Wei Wei.
\newblock Persistent spectral graph.
\newblock {\em International Journal for Numerical Methods in Biomedical Engineering}, 36(9):e3376, 2020.

\bibitem{chen2021evolutionary}
Jiahui Chen, Rundong Zhao, Yiying Tong, and Guo-Wei Wei.
\newblock Evolutionary de {Rham}-{Hodge} method.
\newblock {\em Discrete and continuous dynamical systems. Series B}, 26(7):3785, 2021.

\bibitem{wang2023persistent}
Rui Wang and Guo-Wei Wei.
\newblock Persistent path {L}aplacian.
\newblock {\em Foundations of Data Science}, 5(1):26, 2023.

\bibitem{jones2025persistent}
Benjamin Jones and Guo-Wei Wei.
\newblock Persistent directed flag {L}aplacian.
\newblock {\em Foundations of Data Science}, 7(3):737, 2025.

\bibitem{wei2025persistent2}
Xiaoqi Wei and Guo-Wei Wei.
\newblock Persistent topological {Laplacians}--a survey.
\newblock {\em Mathematics}, 13(2):208, 2025.

\bibitem{wei2025persistent}
Xiaoqi Wei and Guo-Wei Wei.
\newblock Persistent sheaf {Laplacians}.
\newblock {\em Foundations of Data Science}, 7(2):446--463, 2025.

\bibitem{bramer2020atom}
David Bramer and Guo-Wei Wei.
\newblock Atom-specific persistent homology and its application to protein flexibility analysis.
\newblock {\em Computational and Mathematical Biophysics}, 8(1):1--35, 2020.

\bibitem{xia2023persistent}
Kelin Xia, Xiang Liu, and JunJie Wee.
\newblock {\em Persistent Homology for {RNA} Data Analysis}, pages 211--229.
\newblock Springer US, New York, NY, 2023.

\bibitem{pun2020weighted}
Chi~Seng Pun, Brandon Yung~Sin Yong, and Kelin Xia.
\newblock Weighted-persistent-homology-based machine learning for {RNA} flexibility analysis.
\newblock {\em PloS one}, 15(8):e0237747, 2020.

\bibitem{su2025topological}
Zhe Su, Xiang Liu, Layal~Bou Hamdan, Vasileios Maroulas, Jie Wu, Gunnar Carlsson, and Guo-Wei Wei.
\newblock Topological data analysis and topological deep learning beyond persistent homology--a review.
\newblock {\em arXiv preprint arXiv:2507.19504}, 2025.

\bibitem{liu2025}
Xiang Liu, Junjie Wee, and Guo-Wei Wei.
\newblock Topological machine learning for protein-nucleic acid binding affinity changes upon mutation.
\newblock {\em Machine Learning: Science and Technology}, accepted 2025.

\bibitem{harini2025}
Kannan Harini, Masakazu Sekijima, and M.~Michael Gromiha.
\newblock {PRA}-{M}ut{P}red: Predicting the effect of point mutations in protein–{RNA} complexes using structural features.
\newblock {\em Journal of Chemical Information and Modeling}, 65(3):1605--1614, 2025.
\newblock PMID: 39847079.

\bibitem{humphrey1996vmd}
William Humphrey, Andrew Dalke, and Klaus Schulten.
\newblock {VMD}: Visual molecular dynamics.
\newblock {\em Journal of Molecular Graphics}, 14(1):33--38, 1996.

\bibitem{scott2007ribozymes}
William~G Scott.
\newblock Ribozymes.
\newblock {\em Current Opinion in Structural Biology}, 17(3):280--286, 2007.

\bibitem{cech1994representation}
Thomas~R. Cech, Simon~H. Damberger, and Robin~R. Gutell.
\newblock Representation of the secondary and tertiary structure of group {I} introns.
\newblock {\em Nature Structural Biology}, 1(5):273--280, 1994.

\bibitem{ohuchi2002modular}
Shoji~J. Ohuchi, Yoshiya Ikawa, Hideaki Shiraishi, and Tan Inoue.
\newblock Modular engineering of a {Group I} intron ribozyme.
\newblock {\em Nucleic Acids Research}, 30(15):3473--3480, 08 2002.

\bibitem{rich1976transfer}
Alexander Rich and Uttam~L. RajBhandary.
\newblock Transfer {RNA}: Molecular structure, sequence, and properties.
\newblock {\em Annual Review of Biochemistry}, 45:805--860, 1976.

\bibitem{berg2021transfer}
Matthew~D. Berg and Christopher~J. Brandl.
\newblock Transfer {RNA}s: diversity in form and function.
\newblock {\em RNA Biology}, 18(3):316--339, 2021.
\newblock PMID: 32900285.

\bibitem{hayes2023}
Nicole Hayes, Ekaterina Merkurjev, and Guo-Wei Wei.
\newblock Integrating transformer and autoencoder techniques with spectral graph algorithms for the prediction of scarcely labeled molecular data.
\newblock {\em Computers in Biology and Medicine}, 153:106479, 2023.

\bibitem{russold2022persistent}
Florian Russold.
\newblock Persistent sheaf cohomology.
\newblock {\em arXiv preprint}, 2022.
\newblock \mbox{doi}:\url{10.48550/arXiv.2204.13446} (accessed 2023-10-01).

\end{thebibliography}
% can add multiple bib files in separate lines
\bibliographystyle{unsrt}

%\newpage

\appendix

\section{Appendix: Additional Tables}
\label{sec:appendix}

\begin{longtable}[c]{c|cc|cc|cc}
\hline
\multirow{2}{*}{PDB ID} & \multicolumn{2}{c|}{M1 Model}  & \multicolumn{2}{c|}{M2 Model} & \multicolumn{2}{c}{M3 Model} \\
 & PCC & \# of Atoms & PCC & \# of Atoms & PCC & \# of Atoms \\
\endfirsthead
\endhead
\hline
1ASY & 0.647 & 1114 & 0.651 & 1248 & 0.657 & 1382 \\
1C9S & 0.785 & 1583 & 0.804 & 1638 & 0.718 & 1693 \\
1DFU & 0.69 & 130 & 0.744 & 168 & 0.805 & 206 \\
1FEU & 0.478 & 450 & 0.528 & 530 & 0.609 & 610 \\
1GTF & 0.594 & 1571 & 0.577 & 1615 & 0.567 & 1659 \\
1JBR & 0.71 & 356 & 0.695 & 417 & 0.706 & 478 \\
1JBS & 0.637 & 332 & 0.584 & 366 & 0.552 & 400 \\
1K8W & 0.547 & 324 & 0.544 & 345 & 0.562 & 366 \\
1QFQ & 0.774 & 50 & 0.59 & 65 & 0.717 & 80 \\
1SI2 & 0.843 & 128 & 0.719 & 137 & 0.721 & 146 \\
1TTT & 0.628 & 1401 & 0.624 & 1587 & 0.57 & 1773 \\
1U0B & 0.798 & 535 & 0.845 & 609 & 0.814 & 683 \\
1URN & 0.813 & 345 & 0.839 & 401 & 0.831 & 456 \\
1UTD & 0.515 & 1593 & 0.544 & 1628 & 0.516 & 1663 \\
1WNE & 0.62 & 487 & 0.604 & 500 & 0.575 & 513 \\
1YTU & 0.687 & 836 & 0.721 & 855 & 0.741 & 874 \\
1YVP & 0.671 & 1101 & 0.657 & 1145 & 0.648 & 1189 \\
1ZDI & 0.718 & 415 & 0.81 & 444 & 0.712 & 473 \\
1ZH5 & 0.637 & 377 & 0.624 & 395 & 0.537 & 413 \\
2BX2 & 0.329 & 514 & 0.39 & 528 & 0.454 & 542 \\
2C4R & 0.67 & 501 & 0.653 & 511 & 0.665 & 521 \\
2E9T & 0.524 & 974 & 0.597 & 1002 & 0.555 & 1030 \\
2G4B & 0.595 & 179 & 0.66 & 186 & 0.693 & 193 \\
2I91 & 0.624 & 1087 & 0.662 & 1132 & 0.65 & 1177 \\
2IX1 & 0.645 & 656 & 0.625 & 669 & 0.582 & 682 \\
2LA5 & 0.675 & 52 & 0.388 & 88 & 0.289 & 124 \\
2PJP & 0.758 & 144 & 0.794 & 167 & 0.793 & 190 \\
2X1A & 0.711 & 88 & 0.75 & 89 & 0.689 & 90 \\
2XFM & 0.836 & 126 & 0.822 & 133 & 0.777 & 140 \\
2XGJ & 0.541 & 1750 & 0.54 & 1760 & 0.536 & 1770 \\
2XS2 & 0.704 & 93 & 0.704 & 99 & 0.82 & 105 \\
2Y8W & 0.664 & 234 & 0.725 & 254 & 0.779 & 274 \\
2YH1 & 0.65 & 203 & 0.629 & 212 & 0.584 & 221 \\
2ZI0 & 0.796 & 153 & 0.833 & 191 & 0.816 & 229 \\
2ZKO & 0.9 & 182 & 0.792 & 224 & 0.867 & 266 \\
2ZZM & 0.693 & 413 & 0.698 & 497 & 0.791 & 581 \\
2ZZN & 0.486 & 813 & 0.49 & 955 & 0.497 & 1097 \\
3AM1 & 0.587 & 317 & 0.516 & 398 & 0.549 & 479 \\
3EQT & 0.409 & 283 & 0.427 & 299 & 0.466 & 315 \\
3GIB & 0.743 & 200 & 0.734 & 209 & 0.685 & 218 \\
3GPQ & 0.545 & 622 & 0.561 & 626 & 0.572 & 629 \\
3K49 & 0.772 & 1086 & 0.759 & 1116 & 0.731 & 1146 \\
3K5Q & 0.73 & 409 & 0.723 & 418 & 0.716 & 427 \\
3K5Y & 0.725 & 408 & 0.69 & 417 & 0.663 & 426 \\
3L25 & 0.78 & 506 & 0.76 & 522 & 0.785 & 538 \\
3MOJ & 0.678 & 144 & 0.682 & 213 & 0.718 & 282 \\
3NCU & 0.818 & 262 & 0.862 & 284 & 0.87 & 306 \\
3OL6 & 0.633 & 1984 & 0.627 & 2124 & 0.669 & 2264 \\
3QGB & 0.701 & 408 & 0.673 & 417 & 0.667 & 426 \\
3QGC & 0.665 & 408 & 0.651 & 417 & 0.644 & 426 \\
3QSU & 0.651 & 859 & 0.636 & 867 & 0.645 & 875 \\
3RW6 & 0.654 & 609 & 0.575 & 729 & 0.688 & 849 \\
3SN2 & 0.562 & 876 & 0.571 & 905 & 0.544 & 934 \\
3U4M & 0.462 & 308 & 0.598 & 388 & 0.718 & 468 \\
3UZS & 0.607 & 1032 & 0.635 & 1052 & 0.649 & 1072 \\
3V71 & 0.829 & 371 & 0.813 & 378 & 0.829 & 385 \\
3V74 & 0.584 & 405 & 0.571 & 416 & 0.566 & 427 \\
3VYX & 0.666 & 115 & 0.648 & 135 & 0.655 & 155 \\
3VYY & 0.861 & 215 & 0.896 & 265 & 0.91 & 315 \\
3WBM & 0.646 & 393 & 0.656 & 443 & 0.679 & 493 \\
3ZLA & 0.676 & 1919 & 0.671 & 2007 & 0.616 & 2095 \\
4CIO & 0.954 & 103 & 0.959 & 110 & 0.959 & 117 \\
4ED5 & 0.537 & 355 & 0.548 & 373 & 0.535 & 389 \\
4ERD & 0.639 & 256 & 0.556 & 300 & 0.583 & 344 \\
4HT8 & 0.743 & 374 & 0.736 & 388 & 0.715 & 400 \\
4HT9 & 0.717 & 190 & 0.696 & 199 & 0.692 & 208 \\
4M59 & 0.491 & 1400 & 0.513 & 1432 & 0.506 & 1463 \\
4MDX & 0.723 & 238 & 0.686 & 247 & 0.65 & 256 \\
4NGB & 0.695 & 287 & 0.698 & 299 & 0.686 & 311 \\
4NGD & 0.571 & 292 & 0.538 & 304 & 0.599 & 316 \\
4NHA & 0.682 & 246 & 0.732 & 262 & 0.757 & 278 \\
4NKU & 0.65 & 632 & 0.653 & 636 & 0.649 & 640 \\
4O26 & 0.795 & 562 & 0.785 & 656 & 0.68 & 750 \\
4OE1 & 0.492 & 1400 & 0.509 & 1432 & 0.501 & 1463 \\
4OI0 & 0.749 & 397 & 0.763 & 399 & 0.765 & 401 \\
4OOG & 0.615 & 514 & 0.644 & 548 & 0.702 & 582 \\
4PMW & 0.595 & 1410 & 0.595 & 1438 & 0.576 & 1466 \\
4QEI & 0.638 & 630 & 0.701 & 698 & 0.761 & 766 \\
4QG3 & 0.412 & 308 & 0.457 & 388 & 0.533 & 468 \\
4QVC & 0.745 & 367 & 0.741 & 370 & 0.739 & 373 \\
4QVD & 0.767 & 367 & 0.723 & 371 & 0.704 & 375 \\
4QVI & 0.41 & 309 & 0.481 & 389 & 0.585 & 469 \\
4R3I & 0.753 & 167 & 0.747 & 171 & 0.744 & 175 \\
4R8I & 0.803 & 68 & 0.803 & 68 & 0.803 & 68 \\
4RCJ & 0.798 & 191 & 0.782 & 194 & 0.755 & 197 \\
4TUW & 0.76 & 193 & 0.591 & 247 & 0.538 & 299 \\
4V4F & 0.732 & 221 & 0.599 & 228 & 0.675 & 235 \\
4YVI & 0.604 & 559 & 0.685 & 630 & 0.759 & 701 \\
4Z4C & 0.654 & 830 & 0.656 & 857 & 0.696 & 884 \\
4Z4D & 0.602 & 831 & 0.605 & 859 & 0.619 & 886 \\
5AWH & 0.689 & 1544 & 0.668 & 1616 & 0.682 & 1688 \\
5DET & 0.675 & 193 & 0.717 & 202 & 0.694 & 211 \\
5DNO & 0.785 & 170 & 0.639 & 177 & 0.639 & 184 \\
5E3H & 0.532 & 665 & 0.549 & 689 & 0.605 & 713 \\
5EEU & 0.757 & 1565 & 0.774 & 1609 & 0.682 & 1653 \\
5EIM & 0.723 & 332 & 0.69 & 348 & 0.688 & 364 \\
5ELH & 0.652 & 286 & 0.612 & 291 & 0.64 & 296 \\
5ELK & 0.675 & 126 & 0.682 & 130 & 0.678 & 134 \\
5EN1 & 0.801 & 190 & 0.826 & 197 & 0.826 & 204 \\
5EV1 & 0.735 & 204 & 0.731 & 211 & 0.692 & 218 \\
5F98 & 0.55 & 4021 & 0.591 & 4165 & 0.634 & 4309 \\
5GXH & 0.728 & 678 & 0.715 & 684 & 0.706 & 690 \\
5H1K & 0.697 & 1388 & 0.686 & 1413 & 0.683 & 1438 \\
5H1L & 0.816 & 688 & 0.81 & 695 & 0.807 & 702 \\
5HO4 & 0.726 & 188 & 0.705 & 198 & 0.712 & 206 \\
5IP2 & 0.668 & 702 & 0.655 & 713 & 0.657 & 724 \\
5JBJ & 0.514 & 651 & 0.522 & 675 & 0.548 & 699 \\
5M3H & 0.65 & 2221 & 0.581 & 2249 & 0.664 & 2277 \\
5MPG & 0.434 & 103 & 0.429 & 110 & 0.418 & 117 \\
5MPL & 0.362 & 107 & 0.333 & 113 & 0.379 & 119 \\
5UDZ & 0.715 & 323 & 0.658 & 370 & 0.653 & 416 \\
5W1H & 0.72 & 1342 & 0.713 & 1383 & 0.716 & 1424 \\
5WLH & 0.708 & 1332 & 0.705 & 1374 & 0.719 & 1416 \\
5WWW & 0.712 & 99 & 0.645 & 105 & 0.555 & 111 \\
5WWX & 0.644 & 89 & 0.663 & 94 & 0.671 & 99 \\
5WZH & 0.671 & 543 & 0.664 & 554 & 0.686 & 565 \\
5YTV & 0.849 & 79 & 0.852 & 83 & 0.821 & 87 \\
5ZTM & 0.613 & 380 & 0.633 & 428 & 0.665 & 476 \\
6B14 & 0.797 & 523 & 0.785 & 606 & 0.667 & 689 \\
6CMN & 0.611 & 117 & 0.708 & 144 & 0.738 & 171 \\
6CYT & 0.715 & 669 & 0.759 & 686 & 0.773 & 703 \\
6D12 & 0.928 & 237 & 0.949 & 275 & 0.93 & 313 \\
6KWR & 0.801 & 487 & 0.848 & 521 & 0.844 & 555 \\
6NY5 & 0.721 & 769 & 0.712 & 782 & 0.725 & 795 \\
6SDY & 0.325 & 107 & 0.228 & 141 & 0.323 & 175 \\
6SO9 & 0.735 & 117 & 0.742 & 122 & 0.752 & 127\\
\hline
\caption{Pearson correlation coefficients (PCC) for our B-factor predictions of 126 protein-RNA structures \cite{harini2025} using PSL features. Results are given for the three coarse-grained models M1, M2, and M3 of each structure, and the number of atoms in each model is also included.}
\label{tab:126-all-results}\\
\end{longtable}

\end{document}